\documentclass{aa}
\usepackage{txfonts}
\usepackage{graphicx}
\usepackage{natbib}
\usepackage{amssymb}
\bibpunct{(}{)}{;}{a}{}{,}
\begin{document}
   \title{The abundance of SiS in circumstellar envelopes around AGB stars}
 
   \titlerunning{SiS in circumstellar envelopes around AGB stars}

   \author{F.~L. Sch\"oier\inst{1,2} \and J. Bast\inst{1,3} \and H. Olofsson\inst{1,2} \and M. Lindqvist\inst{2}}

   \offprints{F.~L. Sch\"oier \\ \email{schoier@chalmers.se}}

   \institute{Stockholm Observatory, AlbaNova University Center, SE-106 91 Stockholm, Sweden
   \and Onsala Space Observatory, SE-439 92 Onsala, Sweden
   \and Leiden Observatory, P.O. Box 9513, NL-2300 RA Leiden, The Netherlands}

   \date{Received; accepted}

   \abstract{}{Given their photospheric origin and refractive nature, SiS molecules can provide important constraints on the relative roles of dust condensation and non-equilibrium processes in regulating the chemistry in circumstellar envelopes around evolved stars.}{New SiS multi-transition (sub-)millimetre line observations of a sample of AGB stars with varying photospheric C/O-ratios and mass-loss rates are presented. A combination of low- and high-energy lines are important in constraining the circumstellar distribution of SiS molecules. A detailed radiative transfer modelling of the observed SiS line emission is performed, including the effect of thermal dust grains in the excitation analysis. }{We find that the circumstellar fractional abundance of SiS in these environments has a strong dependence on the photospheric C/O-ratio as expected from chemical models. The carbon stars (C/O\,$>$\,1) have a mean fractional abundance of 3.1\,$\times$\,10$^{-6}$, about an order of magnitude higher than found for the M-type AGB stars (C/O\,$<$\,1) where the mean value is 2.7\,$\times$\,10$^{-7}$. These numbers are in reasonable agreement with photospheric LTE chemical models. SiS appears to behave similar to SiO in terms of photodissociation in the outer part of the circumstellar envelope. In contrast to previous results for the related molecule SiO, there is no strong correlation of the fractional abundance with density in the CSE, as would be the case if freeze-out onto dust grains were important. However, possible time-variability of the line emission in the lower $J$ transitions and the sensitivity of the line emission to abundance gradients in the inner part of the CSE may mask a correlation with the density of the wind. There are indications that the SiS fractional abundance could be significantly higher closer to the star which, at least in the case of M-type AGB stars, would require non-equilibrium chemical processes. 
% For the carbon stars, a steep abundance gradient in the inner wind would favour a scenario where SiS molecules are freezing-out onto dust grains.
}{}

   
   \maketitle
%
%________________________________________________________________

\section{Introduction}
Over the past decade it has become increasingly evident that non-equilibrium chemical processes are important in regulating the chemistry in the circumstellar envelopes (CSEs) found around evolved, mass-losing, low- to intermediate mass stars located on the asymptotic giant branch (AGB). From equilibrium chemistry calculations it is expected that the molecular content of the gas and dust  that composes the CSEs to a large extent is dictated by the C/O-ratio in the photosphere of the star  (see reviews by Glassgold 1999\nocite{Glassgold99} and Millar 2003\nocite{Millar03} and references therein). For M-type (O-rich) AGB stars (C/O$<$1) most of the carbon atoms will be tied up in CO leaving an oxygen dominated chemistry with very low abundances of other carbon bearing molecules such as HCN and CS. For carbon stars (C/O$>$1), on the other hand, very low abundances of oxygen bearing species such as, e.g., H$_2$O and SiO are to be expected. 

However, surprisingly large amounts of oxygen bearing molecules such as, e.g., H$_2$O, H$_2$CO, C$_3$O and SiO have been found in the CSE around the carbon star IRC+10216 \citep{Keady93, Melnick01, Ford04, Hasegawa06, Agundez06, Tenenbaum06, Schoeier06c} and, in the case of SiO,  for a large sample of carbon stars indicating that such `anomalous' chemistries might be a common phenomenon in carbon-rich circumstellar envelopes \citep{Schoeier06a}. Various suggestions to explain the observed molecular abundances have been made including non-equilibrium chemical processes, Fischer-Tropsch catalytic processes, and evaporation of cometary bodies \citep{Bieging00, Melnick01, Willacy04, Cherchneff06, Agundez06}. Similarly, in M-type AGB stars relatively large amounts of carbon bearing molecules such as HCN and CS have been found \citep{Olofsson98b}. Recent interferometric line observations of HCN have also sorted out a long standing debate as to its origin, clearly indicating that it is effectively formed in, or close to, the photosphere also in the case of M-type AGB stars \citep{Marvel05}.

It is not unreasonable to expect that also the grain type setups in the CSEs around mass-losing red giant stars are affected by these various chemical processes. Recently,  \citet{Hoefner07} have suggested that non-equilibrium processes forming both carbon and silicate grains in the winds of M-type AGB stars may help these winds to accelerate to the observed values. Such detailed calculations are currently not possible in hydrodynamical simulations for dust driven winds from M-type AGB stars containing pure silicates \citep{Woitke06, Hoefner07}.

Observations of SiO line emission have been shown to be a useful probe of the formation and evolution of dust grains in CSEs, as
well as CSE dynamics. A major survey
of SiO line emission from AGB stars with varying photospheric C/O-ratios and mass-loss rates 
was performed by \citet{Delgado03b} and  \citep{Schoeier06a}. To much surprise, detailed radiative transfer modelling reveal that the circumstellar fractional abundance of SiO has no apparent correlation with the C/O-ratio of the star. For the carbon stars the fractional abundance can be several orders of magnitude higher than predicted by thermal equilibrium chemistry.
Interestingly,  a trend of decreasing SiO abundance with
increasing mass-loss rate of the star, interpreted as an 
effect of increased adsorption of SiO onto dust grains with increasing mass-loss
rate, i.e., the density of the envelope, was found for both the M-type and carbon star samples. 
These claims have been further corroborated by recent, high quality, interferometric SiO observations of two M-type AGB stars \citep{Schoeier04b} and the carbon star IRC+10216  \citep{Schoeier06c} where also high spectral resolution infra-red observations of ro-vibrational transitions were used. 

Thus, there exist strong indications that the circumstellar SiO line emission carries information on the complex region, both chemically and  dynamically, where the mass loss is initiated, and where the dust formation takes place. However, these conclusions still rest on somewhat loose ground, e.g., the constraints on the SiO abundance distribution are poor, and the relative importance of freeze-out onto dust grains, photodissociation, and circumstellar chemistry is still uncertain. 
Important complementary information may be obtained by observing another Si-bearing species, e.g.,
SiS which has been detected in both oxygen- and carbon-rich envelopes \citep{Lindqvist92, Bujarrabal94, Olofsson98b}. This species differs from SiO in the sense that it can be readily formed though gas-phase reactions in the inner wind
\citep{Scalo80}. Furthermore, it could possibly react differently
from SiO to shocks \citep{Cherchneff06}, dust condensation, and photodissociation. The SiS 
lines are expected to be optically thin, as
opposed to the SiO lines which are generally optically thick also for the lowest 
mass-loss-rate objects, and thus has the potential to probe material even closer to the star.

In this paper we present new (sub-)millimetre line observations of SiS for a sample of AGB stars with varying photospheric C/O-ratios and mass-loss rates. The observations are supplemented by a detailed non-LTE radiative transfer analysis in order to obtain reliable circumstellar SiS fractional abundances. The results are then compared to SiO abundance estimates for the same sample of stars and to predictions from available chemical models.

\begin{table}
\caption{Telescope data.}
\label{efficiencies}
$
\begin{array}{cccccccccccc}
\hline
\noalign{\smallskip}
\multicolumn{1}{c}{{\mathrm{Transition}}} & &
\multicolumn{1}{c}{{\mathrm{Frequency}}} &  &
\multicolumn{1}{c}{{E_{\mathrm{up}}}} &  &
\multicolumn{1}{c}{{\mathrm{Telescope}}}  &&
\multicolumn{1}{c}{\eta_{\mathrm{mb}}}  && 
\multicolumn{1}{c}{\theta_{\mathrm{mb}}} \\ 
& & 
\multicolumn{1}{c}{{\mathrm{[GHz]}}} && 
\multicolumn{1}{c}{{\mathrm{[K]}}} &&  
&& &&
\multicolumn{1}{c}{[\arcsec]} \\
\noalign{\smallskip}
\hline
\noalign{\smallskip}
                        J=5\rightarrow4      &&  \phantom{0}90.772  && \phantom{0}13 && \mathrm{OSO}    && 0.60 && 42 \\
                        J=6\rightarrow5      &&                      108.924  && \phantom{0}18 && \mathrm{OSO}    && 0.50 && 35 \\
                        J=12\rightarrow11 &&                      217.817  && \phantom{0}68 && \mathrm{JCMT}  && 0.70 && 22 \\
                        J=19\rightarrow18 &&                      344.779  &&                     166 && \mathrm{APEX}  && 0.70 && 18 \\
                                                          &&                                      &&                              && \mathrm{JCMT}  && 0.63 && 14 \\
                        J=20\rightarrow19 &&                      362.906  &&                     183 && \mathrm{APEX}  && 0.70 && 17\\

\noalign{\smallskip}
\hline
\end{array}
$
\end{table}

\section{Observations}
\label{sect_obs}
Multi-transition SiS line observations (see Table~\ref{efficiencies}) were performed during 1995--2006 using the Onsala 20\,m telescope\footnote{The Onsala 20\,m telescope is operated by the  Swedish National Facility for Radio Astronomy, Onsala Space observatory at Chalmers University of technology} (OSO), the Swedish-ESO submillimetre telescope\footnote{The SEST was located on La Silla, Chile and operated jointly by the Swedish National Facility for Radio Astronomy and the European Southern Observatory (ESO).} (SEST), the JCMT telescope\footnote{Based on observations obtained with the James Clerk Maxwell Telescope, which is operated by the Joint Astronomy Centre in Hilo, Hawaii on behalf of the parent organizations PPARC in the United Kingdom, the National Research Council of Canada and The Netherlands Organization for Scientific Research}, and in July 2006 using the APEX telescope\footnote{This publication is based on data acquired with the Atacama Pathfinder Experiment (APEX). APEX is a collaboration between the Max-Planck-Institut fŸr Radioastronomie, the European Southern Observatory, and the Onsala Space Observatory}.

The SEST, OSO and JCMT observations were made in a dual beamswitch mode, 
where the source is alternately placed in the signal and the reference
beam, using a beam throw of about $11\arcmin$ (SEST and OSO) or $2\arcmin$ (JCMT). 
This method produces very flat baselines. 
At the APEX 12\,m telescope the observations were carried out using a position-switching mode, with the reference position located +3$\arcmin$ in azimuth. 
The raw spectra are stored in the $T_{\mathrm A}^{\star}$  scale and converted to main-beam brightness temperature using $T_{\mathrm{mb}}$\,=\,$T_{\mathrm
A}^{*}/\eta_{\mathrm{mb}}$. $T_{\mathrm A}^{\star}$ is the
antenna temperature corrected for atmospheric attenuation using the
chopper-wheel method, and $\eta_{\mathrm{mb}}$ is the main-beam
efficiency. Regular pointing checks were made on SiO masers (SEST and OSO) and strong CO sources (JCMT and APEX) and typically found to be consistent with the pointing model within $\approx$\,3$\arcsec$.

   \begin{figure*}
   \centering{   
   \includegraphics[width=15cm]{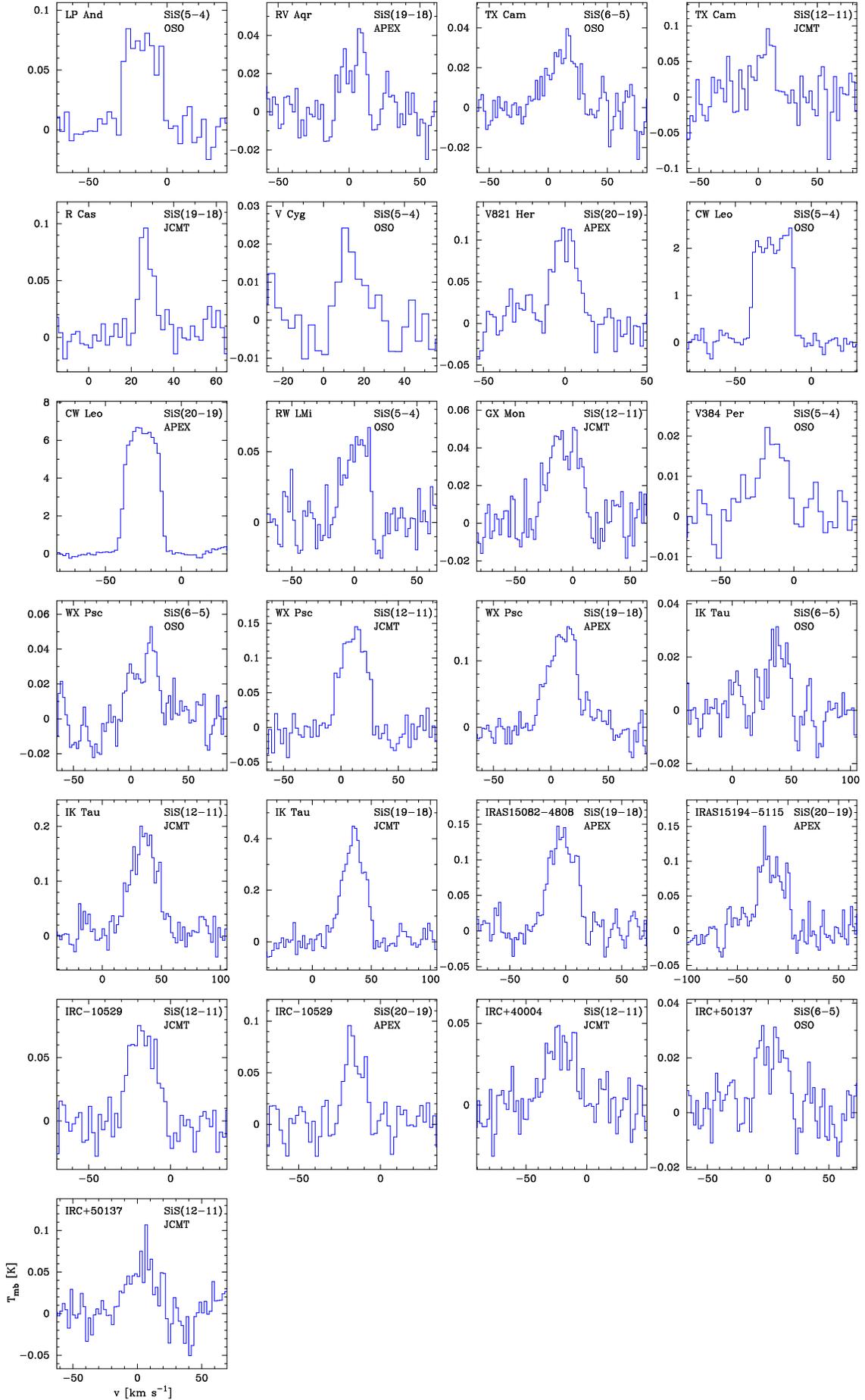}
   \caption{New observations of  SiS ($v$\,$=$\,0, $J$\,$\rightarrow$\,$J-1$)  line emission. The velocity resolution is $\approx$\,2\,$-$\,3\,km\,s$^{-1}$ in all spectra.}
   \label{obs}}
   \end{figure*}
\begin{table*}
\caption{Integrated ($I_{\mathrm{obs}} = \int T_{\mathrm{mb}}\,dv$) line intensities in K\,km\,s$^{-1}$ for the new observations of SiS ($v$\,$=$\,0, $J$\,$\rightarrow$\,$J-1$) line emission.}
\label{intensities}
$
\begin{array}{p{0.15\linewidth}ccccccccc}
\hline
\noalign{\smallskip}
\multicolumn{1}{c}{{\mathrm{Source}}^a} &
\multicolumn{2}{c}{{\mathrm{OSO}}} & &
%\multicolumn{2}{c}{{\mathrm{SEST}}}  &&
\multicolumn{2}{c}{{\mathrm{JCMT}}}  &&
\multicolumn{2}{c}{{\mathrm{APEX}}}   \\ 
\cline{2-3}
%\cline{5-6}
\cline{5-6}
\cline{8-9}
&
\multicolumn{1}{c}{5\rightarrow4} & 
\multicolumn{1}{c}{6\rightarrow5} &&
%\multicolumn{1}{c}{5\rightarrow4}& 
%\multicolumn{1}{c}{6\rightarrow5} &&
\multicolumn{1}{c}{12\rightarrow11} &
\multicolumn{1}{c}{19\rightarrow18} &&
\multicolumn{1}{c}{19\rightarrow18} &
\multicolumn{1}{c}{20\rightarrow19}\\
\noalign{\smallskip}
\hline
\noalign{\smallskip}
{\em carbon stars} \\
\ \object{LP And}       & \phantom{0}1.8\phantom{0} & \cdots && \cdots & \cdots &&  \cdots & \cdots \\
\ \object{RV Aqr}       &  \cdots  &  \cdots  &&  \cdots  &  \cdots && \phantom{00}0.43 & \cdots  \\
%\ \object{U Cam}       & < & \cdots && \cdots & \cdots && \cdots & \cdots \\
%\ \object{S Cep}        &  < & \cdots && \cdots & \cdots && \cdots & \cdots\\
%\ \object{Y CVn}        &  < & \cdots && \cdots & \cdots && \cdots & \cdots\\
\ \object{V Cyg}        & \phantom{0}0.2: & \cdots && \cdots & \cdots &&  \cdots & \cdots \\
%\ \object{R For}         & \\
\ \object{V821 Her}  & \cdots & \cdots && \cdots & \cdots && \cdots & \phantom{00}1.9\phantom{0}\\
\ \object{CW Leo}    & 60\phantom{.00} & \cdots && \cdots & \cdots &&  \cdots & 160\phantom{.00} \\
%\ \object{R Lep}       &\\
\ \object{RW LMi}    & \phantom{0}1.2\phantom{0} & \cdots && \cdots & \cdots &&  \cdots & \cdots \\
\ \object{V384 Per}  & \phantom{0}0.5\phantom{0} & \cdots && \cdots & \cdots &&  \cdots & \cdots \\
%\ \object{W Pic}        & \\
%\ \object{R Vol}         & \\
%\ \object{AFGL 3068}       & \\
%\ \object{IRAS 07454--7112}       & \\
\ \object{IRAS 15082--4808}       &  \cdots & \cdots && \cdots & \cdots && \phantom{00}3.8\phantom{0} & \cdots \\
\ \object{IRAS 15194--5115}       & \cdots & \cdots && \cdots & \cdots && \cdots & \phantom{00}3.0\phantom{0}\\
{\em M-type} \\
\ \object{TX Cam}    &  \cdots & \phantom{0}0.80 && \phantom{0}1.5\phantom{0} & \cdots &&  \cdots & \cdots\\
\ \object{R Cas}        &  \cdots &  \cdots && \cdots & \phantom{0}0.8 && \cdots & \cdots\\
%\ \object{R Crt}          & \\
%\ \object{R Dor}          & \\
%\ \object{R Hya}          & \\
%\ \object{W Hya}          & \\
%\ \object{R Leo}          & \\
\ \object{GX Mon}          &  \cdots & \cdots && \phantom{0}1.4\phantom{0} & \cdots &&  \cdots & \cdots\\
\ \object{WX Psc}       & \cdots & \phantom{0}1.0\phantom{0}  && \phantom{0}4.0\phantom{0} & \cdots && \phantom{00}4.3\phantom{0} & \cdots \\
%\ \object{L$^2$ Pup}          & \\
\ \object{IK Tau}             &  \cdots & \phantom{0}0.40 && \phantom{0}4.7\phantom{0} & \phantom{0}9.2\phantom{0} &&  \cdots & \cdots\\
%\ \object{IRC+10365}          &  \phantom{0}<0.30 & && & && \cdots & \cdots\\
\ \object{IRC--10529}          & \cdots & \cdots && \phantom{0}1.3\phantom{0} & \cdots && \phantom{0}1.0\phantom{0} & \cdots\\
\ \object{IRC+40004}          & \cdots &  \cdots && \phantom{0}1.1\phantom{0} & \cdots && \cdots & \cdots\\
\ \object{IRC+50137}          & \cdots & \phantom{0}0.65  && \phantom{0}1.5\phantom{0} & \cdots && \cdots & \cdots\\
%\ \object{IRC+70066}          &  \phantom{0}<0.30 & \cdots && \cdots & \cdots && \cdots & \cdots\\

\noalign{\smallskip}
\hline
\end{array}
$

$^a$ A colon ($:$) marks a low S/N detection.\\
%$^b$ Improved S/N-ratio compared to previous detection by \citet{Olofsson98b}.
\end{table*}

 The adopted beam efficiencies, together with the FWHM of the main beam ($\theta_{\mathrm{mb}}$), for all telescopes and frequencies are given in Table~\ref{efficiencies}.
The uncertainty in the absolute intensity scale is estimated to be about $\pm 20$\%.  In Table~\ref{efficiencies} also the energy of the upper level involved in the particular transition ($E_{\mathrm{up}}$) is given, ranging from 13\,K for the $J$\,$=$\,5 level up to 183\,K for the  $J$\,$=$\,20 level, illustrating the potential of these multi-transition observations to probe a large radial range of the CSE (see Sect.~\ref{results}).

The data were reduced by removing a low order polynomial baseline and then binned (typicaly to a velocity resolution of about 2\,km\,s$^{-1}$) in order to improve the 
signal-to-noise ratio, using XS\footnote{XS is a package developed by P. Bergman to reduce and analyse a large number of single-dish spectra. It is publically available from {\tt ftp://yggdrasil.oso.chalmers.se}}.

The observed spectra are presented in Fig.~\ref{obs} and velocity-integrated intensities are reported in Table~\ref{intensities}. The intensity
scales are given in main-beam brightness temperature scale ($T_{\mathrm{mb}}$).
In addition to the new data presented here we have also used SiS line intensities ($J$=5$\rightarrow$4 \& $J$=6$\rightarrow$5) reported by  \citet{Bujarrabal94}, \citet{Olofsson98b} and \citet{Woods03}. In total, 11 carbon stars and 8 M-type AGB stars have been detected in SiS line emission. These stars provide our sample and are listed in Table~\ref{sample}.

\begin{table*}
\caption{Model results.}
\label{sample}
$
\begin{array}{p{0.14\linewidth}cccccccccccccccccccccc}
\hline
\noalign{\smallskip}
& 
\multicolumn{8}{c}{\mathrm{SED\ modelling}} & &
\multicolumn{4}{c}{\mathrm{CO\ modelling}} & &
\multicolumn{4}{c}{\mathrm{SiS\ modelling}} \\
\noalign{\smallskip}
\cline{2-9}\cline{11-14}\cline{16-19}
\noalign{\smallskip}
\multicolumn{1}{c}{{\mathrm{Source}}} &
\multicolumn{1}{c}{D}& 
\multicolumn{1}{c}{L_{\star}}&
\multicolumn{1}{c}{T_{\star}}&
\multicolumn{1}{c}{\tau_{10}}&
\multicolumn{1}{c}{T_{\mathrm{c}}}&
\multicolumn{1}{c}{r_{\mathrm{i}}}&
\multicolumn{1}{c}{\chi^2_{\mathrm{red}}} &
\multicolumn{1}{c}{N} & &
\multicolumn{1}{c}{\dot{M}}&
\multicolumn{1}{c}{v_{\mathrm{e}}} &
\multicolumn{1}{c}{\chi^2_{\mathrm{red}}} &
\multicolumn{1}{c}{N} & &
\multicolumn{1}{c}{f_0} & 
\multicolumn{1}{c}{r_{\mathrm{e}}\,^{\mathrm{a}}} &
\multicolumn{1}{c}{\chi^2_{\mathrm{red}}} &
\multicolumn{1}{c}{N} \\
&
\multicolumn{1}{c}{[\mathrm{pc}]}  &
\multicolumn{1}{c}{[\mathrm{L_{\odot}}]} & 
\multicolumn{1}{c}{[\mathrm{K}]} & &
\multicolumn{1}{c}{[\mathrm{K}]} &
\multicolumn{1}{c}{[\mathrm{cm}]} & & & &
\multicolumn{1}{c}{[\mathrm{M_{\odot}\,yr^{-1}}]} &
\multicolumn{1}{c}{[\mathrm{km\,s^{-1}}]} & & 
&&&
\multicolumn{1}{c}{[\mathrm{cm}]} & &  \\
\noalign{\smallskip}
\hline
\noalign{\smallskip}
{\em carbon stars}\\
\ \object{LP And}                       &\phantom{0}630 & \phantom{0}9400 & 2000 & 0.60\phantom{0} & 1100 & 1.8\times10^{14}& 0.8 & 11 && 1.5\times10^{-5} & 13.5 & 0.7 & 7  && \phantom{<:}1.1\times10^{-6}& 2.2\times10^{16} & \phantom{0}2.8 & 3  \\
\ \object{RV Aqr}                      & \phantom{0}670 & \phantom{0}6800 & 2200 & 0.27\phantom{0} & 1300 & 7.6\times10^{13} & 0.8 & 9  & & 2.8\times10^{-6} & 15.0 & 0.3 & 3 && \phantom{<:}1.5\times10^{-6}  & 9.3\times10^{15} & \cdots & 1\\
%\object{UU Aur}                       & \phantom{0}260 & 6900 & 2800 & 0.017 & 1500  &  6.3\times10^{13}& 1.3 & 9 && 2.4\times10^{-7} &10.5& 1.0 & 5  &&  & 3.4\times10^{15} & \cdots & 1 \\
\ \object{U Cam}$^{\mathrm{b}}$ & \phantom{0}340 & \phantom{0}7000  & 2700 & 0.01\phantom{0} &  1500 & 4.4\times10^{13}& 1.5 & 9 && 2.0\times10^{-7} & 11.5 & \cdots & 4 && <7.0\times10^{-5} & 3.0\times10^{15} & \cdots & 1  \\
\ \object{S Cep}                        & \phantom{0}380 & \phantom{0}7300 & 2200 & 0.12\phantom{0} & 1400 & 5.8\times10^{13} & 1.5 & 9 && 1.2\times10^{-6} & 21.5 & 0.9 & 5 && <6.0\times10^{-6}  & 4.8\times10^{15} &\cdots & 1 \\
\ \object{V Cyg}                         & \phantom{0}310 & \phantom{0}6300 & 1900 & 0.08\phantom{0} & 1200 & 8.7\times10^{13}& 0.6 & 8 && 9.0\times10^{-7} & 10.5 & 0.5 & 5 && \phantom{<:}3.5\times10^{-6} & 6.4\times10^{15} & \phantom{0}0.6 & 2 \\
%\object{V1965 Cyg}                &                    1200 & 9800 & 2000 & 0.55\phantom{0} & 1000 & 2.2\times10^{14}& 1.1 & 10 && 1.0\times10^{-5} & 27.0& 2.0 & 3  &&  & 1.4\times10^{16} &  \cdots & 1 \\
%\object{R For}                          &\phantom{0}610 & 5800 & 2000 & 0.25\phantom{0} & 1400 &  5.6\times10^{13}& 2.8 & 9 && 1.1\times10^{-6} & 16.0 & 1.5 & 7 &&  & 5.8\times10^{15} & 0.3 & 4  \\
\ \object{V821 Her}                   & \phantom{0}600 & \phantom{0}7900 & 2200 & 0.45\phantom{0} & 1500 & 8.1\times10^{13}& 2.4 & 10  && 1.8\times10^{-6} & 13.0 & 3.9 & 4  && \phantom{<:}5.5\times10^{-6}  & 8.1\times10^{15} &\cdots & 1 \\
\ \object{CW Leo}                     & \phantom{0}120 & \phantom{0}9600 & 2000 & 0.90\phantom{0} & 1200 & 1.7\times10^{14}& 2.1 & 9 && 1.5\times10^{-5} & 14.0 & 0.5 & 8 && \phantom{<:}1.4\times10^{-6}  & 2.2\times10^{16} & \phantom{0}2.5 & 6 \\
%\object{R Lep}                         & \phantom{0}250 & 4000 & 2200 & 0.06\phantom{0} &  1500 & 4.3\times10^{13}& 0.3 & 9 && 5.0\times10^{-7} & 16.5 & 0.7 & 6 &&  & 4.5\times10^{15} & 0.9 & 4  \\
\ \object{RW LMi}                      & \phantom{0}440 & \phantom{0}9700 & 2000 & 0.50\phantom{0} & 1000 & 2.1\times10^{14}& 1.4 & 11 && 6.0\times10^{-6} & 16.5 & 1.1 & 7 && \phantom{<:}4.0\times10^{-6}  & 1.3\times10^{16} & \phantom{0}0.9 & 3  \\
\ \object{V384 Per}                   & \phantom{0}560 & \phantom{0}8100 & 2000 & 0.25\phantom{0} & 1300 & 1.0\times10^{14}& 1.5 & 11  && 3.5\times10^{-6} & 14.5 & 0.7 & 6 &&  \phantom{<:}2.5\times10^{-6}  & 1.1\times10^{16} & \cdots & 1 \\
%\object{W Pic}                          &   \phantom{0}490  & 4000 & 2500 & 0.015  & 1500  & 4.5\times10^{13} &  3.2  & 8  && 2.3\times10^{-7} & 15.0& 3.4 & 3  &&  & 2.8\times10^{15} & \cdots & 1 \\
\ \object{R Vol}                          &\phantom{0}730 & \phantom{0}6800 & 2000 & 0.30\phantom{0} & 1500 & 6.6\times10^{13}& 1.1 &  9 && 1.7\times10^{-6} & 16.5 & 0.6 & 3 &&  <3.0\times10^{-5}  & 7.0\times10^{15} & \cdots & 1  \\
\ \object{AFGL 3068}                & \phantom{0}980 & \phantom{0}7800 & 2000 & 2.70\phantom{0} & 1100 & 2.5\times10^{14}& 1.8 & 8  && 1.0\times10^{-5} & 13.5 & 0.6 & 4  && \phantom{<:}3.5\times10^{-6} & 1.8\times10^{16} & \phantom{0}6.2 & 3 \\
\ \object{IRAS\,07454--7112}  &\phantom{0}710 & \phantom{0}9000 & 2100 & 0.45\phantom{0} & 1200 & 1.4\times10^{14}&  1.0 & 9 && 5.0\times10^{-6} & 12.5 & 0.1 & 2 && \phantom{<:}6.5\times10^{-6}  & 1.3\times10^{16} & \phantom{0}2.6  & 2  \\
\ \object{IRAS\,15082--4808}  &\phantom{0}640 & \phantom{0}9000 & 2200 & 0.80\phantom{0} & 1100 & 1.9\times10^{14}& 7.0 & 9 && 1.0\times10^{-5} &19.0 & 0.2 & 2  && \phantom{<:}3.0\times10^{-6} & 1.5\times10^{16} &  \phantom{0}7.4 & 3\\
\ \object{IRAS\,15194--5115}  & \phantom{0}500 & \phantom{0}8800 & 2400 & 0.55\phantom{0} & 1200 & 1.5\times10^{14}& 0.4 &  9 && 9.0\times10^{-6} & 21.0 & 0.9 & 4   && \phantom{<:}2.0\times10^{-6}  & 1.4\times10^{16} & 15.8 & 3 \\
{\em M-type}\\
\ \object{TX Cam}                     &\phantom{0}380 & \phantom{0}8600 & 2600 & 1.00\phantom{0} & 1300 & 1.0\times10^{14}& 1.1 & 8 && 1.0\times10^{-5} & 18.0 & 1.0 & 5  && \phantom{<:}4.0\times10^{-7} & 1.6\times10^{16} & 2.0 & 4  \\
\ \object{R Cas}                        &\phantom{0}110 & \phantom{0}3500 & 2800 & 0.06\phantom{0} & \phantom{0}900 & 1.4\times10^{14}& 0.3 & 8 && 5.0\times10^{-7} & 10.0 & 2.0 & 4  && \phantom{<:}4.0\times10^{-7} & 5.0\times10^{15} & \cdots & 1  \\
\ \object{GX Mon}                     &\phantom{0}550 & \phantom{0}8200 & 2600 & 2.00\phantom{0} & \phantom{0}900 & 1.1\times10^{14}& 5.0 & 9 && 2.0\times10^{-5} & 18.5 & 2.0 & 4  && \phantom{<:}1.1\times10^{-7} & 2.2\times10^{16} & 3.7 & 3  \\
\ \object{WX Psc}                     &\phantom{0}700 & 10300 & 1800 & 3.00\phantom{0} & 1000 & 1.1\times10^{14}& 1.3 & 9 && 4.0\times10^{-5} & 18.5 & 2.0 & 4  && \phantom{<:}1.7\times10^{-7} & 3.0\times10^{16} & 7.2 & 4  \\
\ \object{IK Tau}                        &\phantom{0}260 & \phantom{0}7700 & 2400 & 2.00\phantom{0} & 1500 & 7.8\times10^{13}& 1.0 & 9 && 1.0\times10^{-5} & 18.0 & 0.7 & 4  && \phantom{<:}1.0\times10^{-7} & 1.6\times10^{16} & 9.9 & 4  \\
\ \object{IRC--10529}              &\phantom{0}620 & 10600 & 2000 & 3.50\phantom{0} & 1100 & 1.8\times10^{14}& 2.2 & 8 && 3.0\times10^{-5} & 13.0 & 4.0 & 5  && \phantom{<:}1.0\times10^{-7} & 3.1\times10^{16} & 0.1 & 3  \\
\ \object{IRC+40004}              &\phantom{0}680 & 11800 & 2000 & 0.90\phantom{0} & 1100 & 1.5\times10^{14}& 0.9 & 7 && 1.5\times10^{-5} & 17.5 & 0.9 & 2  && \phantom{<:}3.5\times10^{-7} & 1.9\times10^{16} & \cdots & 1  \\
\ \object{IRC+50137}              & 1200 & \phantom{0}9800 & 2000 & 3.50\phantom{0} & 1100 & 1.7\times10^{14}& 1.2 & 7 && 3.0\times10^{-5} & 16.5 & 1.0 & 2  && \phantom{<:}5.0\times10^{-7} & 2.0\times10^{16} & 0.9 & 3  \\

\noalign{\smallskip}
\hline
\end{array}
$
\noindent
$^{\mathrm{a}}$  The SiS envelope size is determined from Eq.~\ref{eq_size}.\\
$^{\mathrm{b}}$  U Cam has a detached shell that complicates the analysis of the present-day mass-loss characteristics \citep[for details see][]{Schoeier05b}.\\
\noindent
%A colon (:) marks an uncertain abundance estimate.
\end{table*}

\section{SiS excitation analysis}
\label{sect_model}
\subsection{Radiative transfer model}
The CSEs are assumed to be spherically symmetric, produced by a constant mass loss rate ($\dot{M}$), and to expand at a constant velocity ($v_{\mathrm e}$). 
%From the continuity equation the density in the wind, with radial distance from the star $r$, can be found.
%
%\begin{equation}
%\rho_{\mathrm{H_2}} = \frac{\dot{M}}{4\pi r^2 v_{\mathrm e}},
%\end{equation}
%
%where $\dot{M}$ is the gas mass-loss rate. 
The SiS excitation analysis is performed using a detailed non-LTE radiative transfer code, based on the Monte Carlo method and is described in more detail in \citet{Schoeier01}.  The code has been extensively tested and benchmarked against a wide variety of molecular-line radiative
transfer codes in \citet{Zadelhoff02}.  

The circumstellar physical properties of the gas, such as the density, temperature, and kinematic structures, of the sample stars is based on radiative transfer modelling of multi-transitional (sub-)millimetre CO line observations. The CO data used in the analysis have been presented in \citet{Schoeier01}, \citet{Olofsson02}, \citet{Delgado03b}, and \citet{Schoeier06a}. The kinetic temperature structure is obtained in a self-consistent manner from solving the energy balance equation, where the CO line cooling is directly obtained from the excitation analysis. The local line width is assumed to be described by a Gaussian and is made up of a micro-turbulent component with a Doppler width of 1.0\,km\,s$^{-1}$ ($e$-folding radius) and a thermal component that is directly  calculated from the derived kinetic temperature structure.

The excitation analysis includes radiative excitation through the first vibrationally excited ($v$\,=\,1) state for CO at 4.6\,$\mu$m and SiS at 13\,$\mu$m. Both the central star (approximated by a blackbody) and thermal dust grains distributed in the CSE can provide sufficient radiation fields that start to populate the $v$\,=\,1 state, and thereby affecting the excitation in the ground vibrational state. Relevant molecular data are summarized in \citet{Schoeier05a} and are made publicly available through the {\em Leiden Atomic and Molecular Database} (LAMDA){\footnote{\tt http://www.strw.leidenuniv.nl/$\sim$moldata}}. The collisional rate coefficients have been extended to include more energy levels as well as extrapolated in temperature, as described in \citet{Schoeier05a}. An ortho-to-para ratio of 3.0 was adopted when weighting together collisional rate coefficients for CO in collisions with ortho-H$_2$ and para-H$_2$. For SiS the same set of collisional rate coefficients as calculated for SiO has been adopted \citep{Schoeier05a}.

The majority of sources in this study are intermediate- to high-mass-loss-rate objects where thermal dust emission provides the main source of infrared photons which excite the $v$\,=\,1 state. The addition of a dust component in the Monte Carlo scheme is straightforward as described in \citet{Schoeier02b}. The dust-temperature structure and dust-density profile are obtained from detailed radiative transfer modelling using {\em Dusty} \citep{Ivezic97}. In the modelling, where the SED provides the observational constraint, the dust optical depth specified at 10\,$\mu$m, $\tau_{10}$, and 
the dust condensation temperature, $T_{\mathrm{c}}$\,$=$\,$T_{\mathrm{d}} (r_{\mathrm{i}})$, are the adjustable parameters in the $\chi^2$-analysis. The effective stellar blackbody temperature, $T_{\star}$, is usually less well constrained for these intemediate- to high mass-loss-rate objects. The SED typically consists of {\em{JHKLM}} photometric data \citep[][ and Kerschbaum priv. com.]{Kerschbaum99b}, IRAS fluxes, and in some cases sub-millimetre data \citep{Groenewegen93}. The total luminosity ($L_{\star}$) of the source is obtained from the period-luminosity relation of \citet{Groenewegen96} and the distance is obtained from the SED fitting. For the sources where good Hipparcos parallaxes exist the corresponding distance has been adopted and the luminosity is then obtained from the SED modelling. Amorphous carbon dust grains 
with the optical constants given in \citet{Suh00} are adopted for the carbon stars while astronomical silicates are adopted for the M-type AGB stars. For simplicity, the 
dust grains are assumed to be of the same size (a radius, $a_{\mathrm{d}}$, of 0.1\,$\mu$m), and the same mass density ($\rho_{\mathrm{s}}$\,=\,2.0\,g\,cm$^{-3}$). 
The corresponding dust opacities, $\kappa_{\nu}$, were then calculated from the optical constants and the individual grain properties using standard Mie theory \citep{Bohren83}. 

The parameters obtained from the CO excitation analysis and the dust modelling describing the physical properties of the circumstellar  envelopes are reported in Table~\ref{sample}.
 
\subsection{SiS abundance distribution}
The SiS fractional abundance distribution is assumed to be described by a Gaussian
\begin{equation}
\label{eq_distr}
f(r) = f_0\, \exp \left(-\left(\frac{r}{r_{\mathrm e}}\right)^2 \right),
\end{equation}
where $f$\,$=$\,$n\mathrm{(SiS)}/n\mathrm{(H_2)}$, i.e., the ratio of the number density of SiS molecules to that of H$_2$ molecules. 
In circumstellar envelopes such as these H is expected to be mainly in molecular form. In Sect.~\ref{results} we will discuss deviations from a Gaussian fractional abundance distribution.

   \begin{figure}
   \centering{   
   \includegraphics[width=7cm]{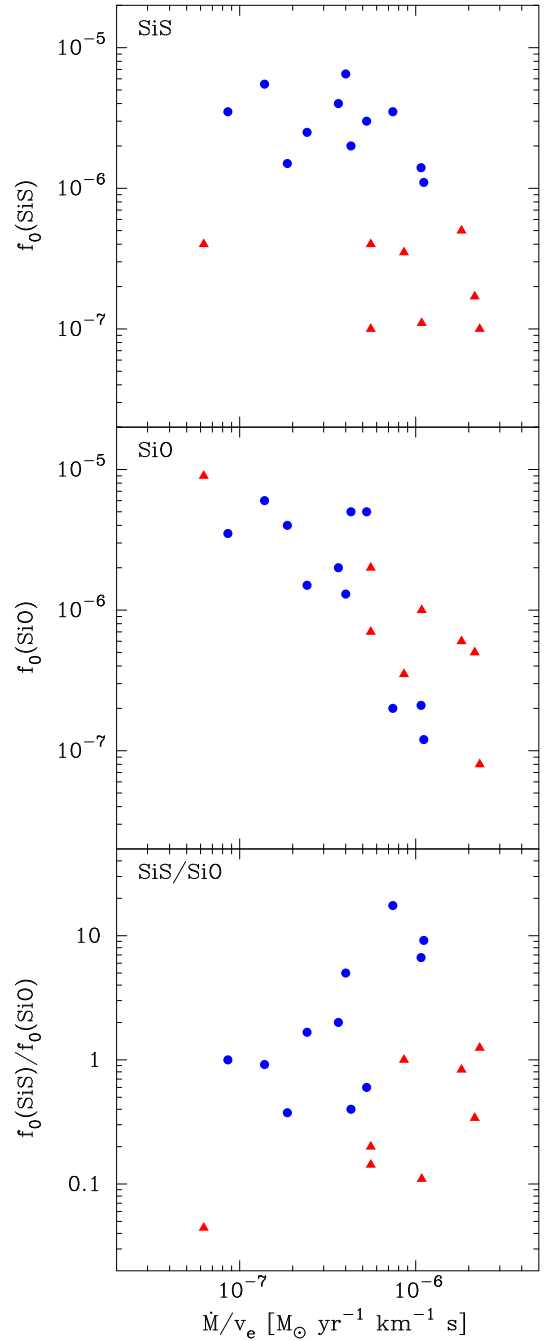}
   \caption{{\em Upper panel -} SiS fractional abundance ($f_0$) obtained from an excitation analysis,  as a function of a density measure ($\dot{M}$/$v_{\mathrm{e}}$), for carbon stars (filled circles) and M-type (O-rich) AGB stars (filled triangles). 
  {\em Middle panel - } Same as above for SiO for the same sample of stars (Sch\"oier et al.\ 2006b; Sch\"oier et al., in prep.). {\em Lower panel - } The SiS/SiO-abundance ratio.}
   \label{sio_abundance}}
   \end{figure}

The best fit model is found by minimizing the total $\chi^2$ defined as
\begin{equation}
\label{chi2_sum}
\chi^2_{\mathrm{tot}} = \sum^N_{i=1} \left [ \frac{(I_{\mathrm{mod}}-I_{\mathrm{obs}})}{\sigma}\right ]^2, 
\end{equation} 
where $I$ is the integrated line intensity and $\sigma$ the uncertainty in the measured 
value (usually dominated by the calibration uncertainty of $\pm$20\%), and the summation is done over
$N$ independent observations. 

The size of the SiS molecular envelopes are assumed to be the same as those for SiO, i.e., it is assumed to behave in the same way to photodissociation \citep[e.g.,][]{Wirsich94}. 
\citet{Delgado03b} and \citet{Schoeier06a} found that a scaling law 
\begin{equation}
\label{eq_size}
\log r_{\mathrm{e}} = 19.2 + 0.48 \log \left( \frac{\dot{M}}{v_{\mathrm e}} \right),
\end{equation}
where $\dot{M}$ is the mass-loss rate and $v_{\mathrm e}$ the expansion velocity of the wind, provide a good fit in the case of SiO for a large sample of M-type and carbon stars. Using this scaling law only one free parameter remains, the fractional abundance of SiS ($f_0$). The adopted sizes using Eq.~\ref{eq_size} are reported in Table~\ref{sample}. In Sect.~\ref{envelope_size} the validity of this approximation is tested for the two carbon stars \object{CW Leo} and \object{RW LMi} which are the only  sources where high angular interferometric observations of SiS line emission exist.

\section{Results}
\label{results}
\subsection{SiS abundances}
The derived abundances for the sample of 19 AGB stars where SiS line emission has been detected are reported in Table~\ref{sample} and Fig.~\ref{sio_abundance}. It is found that the fractional abundance of SiS varies substantially between the carbon star (filled circles in Fig.~\ref{sio_abundance}) and M-type samples (filled squares in Fig.~\ref{sio_abundance}). The mean fractional abundance of SiS in carbon stars of 3.1\,$\times$\,10$^{-6}$ (11 objects) is about an order of magnitude higher than found for the M-type AGB stars (on average 2.7\,$\times$\,10$^{-7}$; 8 objects).

There is a weak trend that the fractional SiS abundance decreases as the density of the wind ($\dot{M}$/$v_{\mathrm e}$) increases. The Pearson correlation coefficients are $r$\,$=$\,$-0.36$ for the carbon stars and $r$\,$=$\,$-0.33$ for the M-type AGB stars. In comparison, for the same sample of sources, the derived SiO abundances have much stronger correlations with $r$\,$=$\,$-0.60$ for the carbon stars and $r$\,$=$\,$-0.64$ for the M-type AGB stars. The correlation in the case of SiO is further increased for the larger sample of sources presented in \citet{Delgado03b} and \citet{Schoeier06a} and interpreted as increased importance of adsorption of SiO molecules onto dust grains as the density in the wind increases.

The fractional abundances of SiS derived agrees well, typically within a factor of two, with those obtained by \citet{Woods03} using a simple excitation model, for the five high mass-loss rate objects in common. The abundances derived in the present analysis are usually higher than the ones reported by \citet{Woods03} possibly a result of their adopted excitation temperature being too high. \citet{Bujarrabal94} find SiS fractional abundances that on average are higher for the carbon stars than for the M-type AGB stars, supporting the findings in this paper. However, they only detect SiS emission for two M-type AGB stars and four carbon stars. The fractional abundances that \citet{Bujarrabal94}, through a simple analysis, derive agree within a factor of five with the values given in Table~\ref{sample} for the 6 sources in common.

   \begin{figure*}
   \centering{   
   \includegraphics[width=17cm]{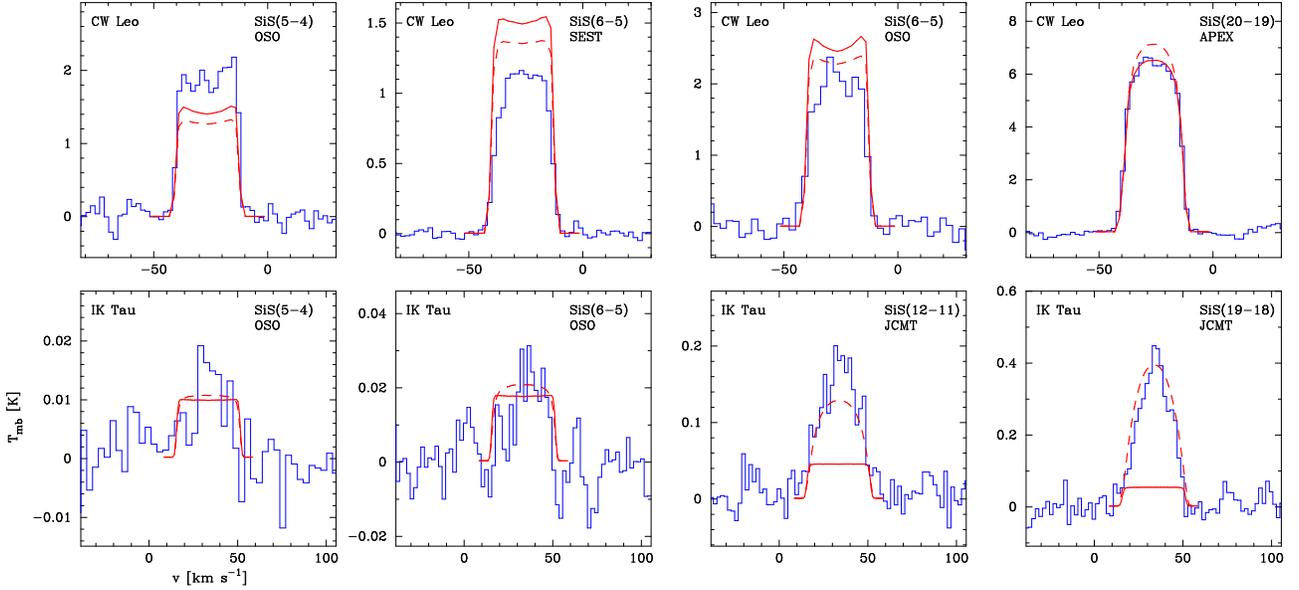}
   \caption{Best-fit models (solid lines; parameters given in Table~\ref{sample}) for the carbon star \object{CW~Leo} and the M-type AGB star \object{IK~Tau} overlayed on observed spectra (histograms). Also models with an additional compact, high SiS abundance,  pre-condensation component are shown.  (dashed line; see text for details).}
   \label{model}}
   \end{figure*}

The fits to the SiS multi-transitional data are generally not as good as for CO or SiO. As examples we show the best-fit
models for the carbon star \object{CW~Leo} and the M-type AGB star  \object{IK~Tau} in Fig.~\ref{model}. For \object{CW~Leo} a reasonably good fit to the observations, both in term of the velocity-integrated line intensity ($\chi^2_{\mathrm{red}}$\,$=$\,2.5) as well as line profile, is found. In contrast, for \object{IK Tau} both the line intensities ($\chi^2_{\mathrm{red}}$\,$=$\,9.9) and line profiles poorly fit the observations. In particular for the high-$J$ transitions, where the signal-to-noise is high, the models predict line profiles that are clearly much more flat-topped than what is actually observed. In Sect.~\ref{high_chi2} possible causes for the generally high $\chi^2$-values will be discussed, including a compact, high fractional abundance region close to the star which will naturally produce parabolic line profiles.

\begin{figure}
      \centering{   
   \includegraphics[width=7cm]{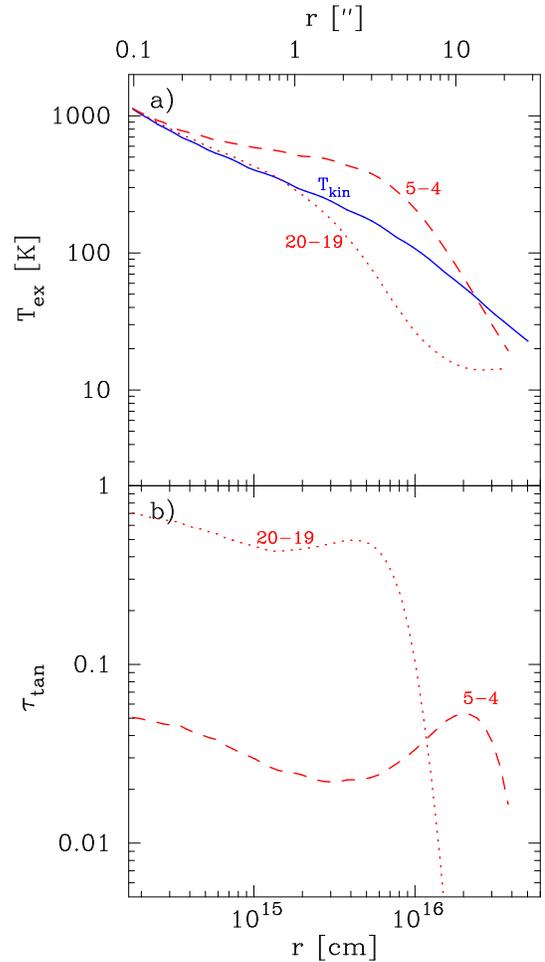}
   \caption{Results for the best-fit model for \object{CW Leo} from the excitation analysis.  {\bf a)} The kinetic temperature of the gas particles as derived from the CO modelling is shown by the solid line. The dotted and dashed lines gives the excitation temperature of the SiS $J$\,=\,5\,$\rightarrow$\,4 and $J$\,=\,20\,$\rightarrow$\,19 transitions, respectively. {\bf b)} The dotted and dashed lines gives the tangential optical depth ($\tau_{\mathrm{tan}}$) at the line center of the SiS $J$\,=\,5\,$\rightarrow$\,4 and $J$\,=\,20\,$\rightarrow$\,19 transitions, respectively.}
   \label{excitation_fig}}
\end{figure}

Results from the SiS excitation analysis in the case of the carbon star \object{CW Leo} are shown in Fig.~\ref{excitation_fig}, where the excitation temperature, $T_{\mathrm{ex}}$, and the tangential optical depth at the line center, $\tau_{\mathrm{tan}}$, are plotted as functions of radial distance from the star for the  $J$\,$=$\,5\,$\rightarrow$\,4 and $J$\,$=$\,20\,$\rightarrow$\,19 transitions. It is clear that both lines are formed under non-LTE  conditions as $T_{\mathrm{ex}}$ is far from the kinetic temperature of the gas ($T_{\mathrm{kin}}$, also shown in Fig.~\ref{excitation_fig}) where  $\tau_{\mathrm{tan}}$ is peaking. Both lines are also optically thin (near the line centre) even in this high mass loss rate object. 

\subsection{Sensitivity analysis}
\label{tests}
In the SiS excitation analysis many parameters have been fixed to values obtained from modelling of other molecules (such as CO and SiO) or thermal dust emission. These parameters, such as, e.g., the mass loss rate (density structure) and temperature structure have some uncertainties. In Table~\ref{table_rovib} we report results of an extensive sensitivity analysis where the values adopted for several of the fixed parameters have been varied within reasonable limits, typically $\pm$\,30\%  for the high mass loss rate carbon star \object{CW Leo} and the low mass loss rate M-type AGB star \object{R Cas}. The calculations were performed for a 15\,m telescope (such as the SEST or JCMT) and the velocity-integrated line intensities for selected transitions from each new model were compared to those from the best fit model presented in Table~\ref{sample}. 

Varying the mass loss rate ($\dot{M}$), i.e. the H$_2$ density scale, by $\pm$\,30\% results in an almost linear response in the line intensities indicating that the emission is more or less optically thin (see also Fig.~\ref{excitation_fig}). These changes only marginally affect the line ratios by at most some 10\%. A similar result is obtained from varying the fractional SiS abundance ($f_0$). The line intensities are significantly affected by changes in the SiS envelope size. As expected it is found that the lower lying (in frequency) $J$\,$=$\,5\,$\rightarrow$\,4 and $J$\,$=$\,6\,$\rightarrow$\,5 transitions, which probe material at larger radial distances from the star (see Fig.~\ref{excitation_fig}), are most sensitive to the adopted SiS envelope size. The $J$\,$=$\,19\,$\rightarrow$\,18 and $J$\,$=$\,20\,$\rightarrow$\,19 line emission are hardly affected at all. This means that line intensity ratios, and hence the quality of the model in fitting multi-transition observations,  will be sensitive to the adopted envelope size, more so than changes in the fractional SiS abundance.

Changing the absolute scale of the kinetic temperature structure, obtained in a self-consistent manner from the CO analysis, by  $\pm$\,30\% only affects the line intensities by a few percent. This indicates that collisional excitation of the SiS lines is not very effective. This is further supported by variation in the external radiation fields. The excitation analysis includes the possibility of populating also the first vibrationally-excited state of SiS through the absorption of 13\,$\mu$m photons, primarily, from dust emission. This can significantly change the derived line intensities of transitions within the ground vibrational state as reported in Table~\ref{table_rovib}. For the high mass loss rate object  \object{CW Leo} the thermal dust emission dominates over the direct stellar radiation field in the excitation.  For the low mass loss rate object \object{R Cas} the radiation fields from the star and circumstellar dust appear to be about equally important. We also note here that the high-$J$ lines are more sensitive to the radiation field than the low-$J$ lines affecting the line intensity ratios.

In an analysis such as this it is very difficult to put absolute errors on the derived fractional abundances of SiS listed in Table~\ref{sample}. However, based on our sensitivity analysis a conservative estimate would be that they are reliable to within a factor of two to three, given that the assumption of a single-Gaussian abundance profile is met. The potential of detecting chemical gradients in the SiS fractional abundance distribution will be further discussed in Sect.~\ref{compact}.

\begin{table}
\caption{Sensitivity tests for a 15\,m telescope$^{\mathrm a}$.}
\label{table_rovib}
$
\begin{array}{p{0.25\linewidth}ccccc}
\hline
\noalign{\smallskip}
&
\multicolumn{1}{c}{5\rightarrow4} & 
\multicolumn{1}{c}{6\rightarrow5} & 
\multicolumn{1}{c}{12\rightarrow11}  &
\multicolumn{1}{c}{19\rightarrow18}  &
\multicolumn{1}{c}{20\rightarrow19}  \\ 

\multicolumn{1}{c}{{\mathrm{Model\ with}}} &
\multicolumn{1}{c}{[\%]} &
\multicolumn{1}{c}{[\%]} &
\multicolumn{1}{c}{[\%]} &
\multicolumn{1}{c}{[\%]} &
\multicolumn{1}{c}{[\%]} \\
\noalign{\smallskip}
\hline
\noalign{\smallskip}
{\em CW Leo} \\
$\dot{M}$\,$+$\,30\% & +28 & +26 & +21 & +22 & +21 \\
$\dot{M}$\,$-$\,30\% & -27 & -27 & -26 & -27 & -27  \\
$f_0$\,$+$\,30\% & +29 & +27 & +21& +19 & +19  \\
$f_0$\,$-$\,30\% &-29 &  -28 & -25 & -24 & -24 \\
$r_{\mathrm{e}}$\,$+$\,30\% & +44 & +35  &  +7 & 0 & 0 \\
$r_{\mathrm{e}}$\,$-$\,30\% & -42 & -39 & -16 & -5 & -5 \\
$T_{\mathrm{kin}}(r)$\,$+$\,30\% & -1 & -1 & +1 & +5 & +5\\
$T_{\mathrm{kin}}(r)$\,$-$\,30\% & +4 & +3 &  -3 & -9 & -10 \\
no star                                              &   0 & 0 & -1 & -1 & -1 \\
no dust                                             &  +5 & -12 & -37 & -37 & -38\\

\noalign{\smallskip}
{\em R Cas} \\
$\dot{M}$\,$+$\,30\% & +23 & +24 & +31 & +32 & +32 \\
$\dot{M}$\,$-$\,30\% & -24 & -25 & -30 & -31 & -31  \\
$f_0$\,$+$\,30\% & +27 & +27 & +27& +26 & +25 \\
$f_0$\,$-$\,30\% &-27 &  -28 & -27 & -26 & -26 \\
$r_{\mathrm{e}}$\,$+$\,30\% & +45 & +37  &  +9 & +2 & +1 \\
$r_{\mathrm{e}}$\,$-$\,30\% & -42 & -39 & -16 & -4 & -4 \\
$T_{\mathrm{kin}}(r)$\,$+$\,30\% & -4 & -3 & +2 & +7 & +8\\
$T_{\mathrm{kin}}(r)$\,$-$\,30\% & +4 & +3 &  -3 & -8 & -9 \\
no star                                              &   -6  & -6 & -11 & -15 & -16 \\
no dust                                             &  +9 & +1 & -14 & -12 & -12\\

\noalign{\smallskip}
\hline
\end{array}
$

\noindent
$^{\mathrm{a}}$ The percentage change in velocity-integrated intensity relative to the best-fit model presented in Table~\ref{sample}.\\
\end{table}
   \begin{figure}
   \centering{   
   \includegraphics[width=6cm, angle=-90]{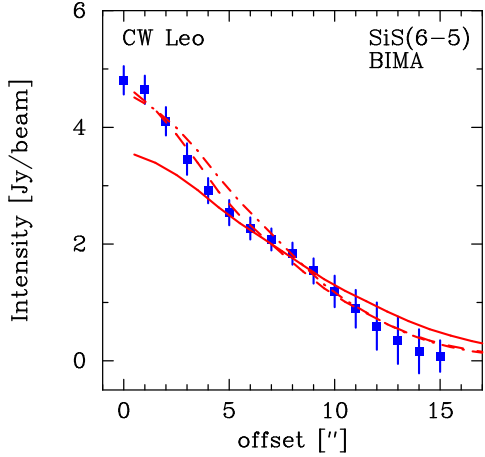}
   \caption{Azimuthally averaged brightness distribution for \object{CW Leo} (\object{IRC+10216}), mean value around the systemic velocity from $-24$ to $-28$\,km\,s$^{-1}$ (LSR), obtained using the BIMA interferometer (data from Bieging \& Tafalla 1993) as a function of radial distance to the star in arcsec. The angular resolution is 8$\arcsec$ (FWHM). The best-fit single-dish model using $r_{\mathrm{e}}$\,$=$\,2.2\,$\times$\,10$^{16}$\,cm and $f_0$\,$=$\,1.5\,$\times$\,10$^{-6}$ is indicated by the solid line. A model (dash-dotted line) with $r_{\mathrm{e}}$\,$=$\,1.7\,$\times$\,10$^{16}$\,cm and $f_0$\,$=$\,2.0\,$\times$\,10$^{-6}$ better reproduces the interferometer data. A still better fit (dashed line) is obtained by introducing a compact SiS component  ($r_{\mathrm{c}}$\,$=$\,5.0\,$\times$\,10$^{14}$\,cm and $f_{\mathrm{c}}$\,$=$\,1.7\,$\times$\,10$^{-5}$) in addition to the more extended envelope with $r_{\mathrm{e}}$\,$=$\,1.8\,$\times$\,10$^{16}$\,cm and $f_0$\,$=$\,1.7\,$\times$\,10$^{-6}$. All models are consistent with available single-dish data within 1$\sigma$.}
   \label{cwleo}}
   \end{figure}
   \begin{figure}
   \centering{   
   \includegraphics[width=7cm, angle=-90]{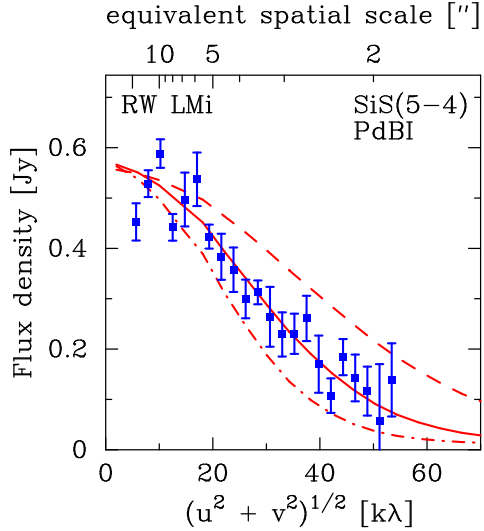}
   \caption{Azimuthally averaged visibility amplitudes (real part) for \object{RW LMi} (\object{CIT6}), mean value around the systemic velocity from $-2$ to $0$\,km\,s$^{-1}$ (LSR), obtained using the Plateu de Bure interferometer (data from Lindqvist et al.\ 2000) as a function of distance to the phase center in k$\lambda$ (also shown on the upper abscissa is the equivalent spatial resolution in arcseconds). The best-fit single-dish model using $r_{\mathrm{e}}$\,$=$\,1.3\,$\times$\,10$^{16}$\,cm and $f_0$\,$=$\,4.5\,$\times$\,10$^{-6}$ is indicated by the solid line. Also shown are models with a 30\% smaller envelope size ($r_{\mathrm{e}}$\,$=$\,9.0\,$\times$\,10$^{15}$\,cm and $f_0$\,$=$\,8.5\,$\times$\,10$^{-6}$; dashed line) and 30\% larger envelope size ($r_{\mathrm{e}}$\,$=$\,1.7\,$\times$\,10$^{16}$\,cm and $f_0$\,$=$\,2.8\,$\times$\,10$^{-6}$; dash-dotted line). All models are consistent with available single-dish data within 1$\sigma$.}
   \label{cit6}}
   \end{figure}

\subsection{What do the high $\chi^2_{\mathrm{red}}$ tell us?}
\label{high_chi2}
The resulting $\chi^2_{\mathrm{red}}$ are, in the majority of the sources, higher than two, indicating relatively poor fits to the multi-transitional SiS data. This is in contrast to, e.g., the excitation analysis of SiO line emission for the same sample of sources, where  $\chi^2_{\mathrm{red}}$\,$\approx$\,$1-2$ in general. There are a number of potential reasons for this that will be addressed below.

\subsubsection{SiS envelope size}
\label{envelope_size}
In the present analysis we simply assume that the SiS circumstellar molecular envelope has the same size as that of SiO, i.e., it is assumed that these two molecules behave in the same way to photodissociation. This appears to be a reasonable first approximation \citep[e.g.,][]{Wirsich94}. However, the sensitivity tests performed in Sect.~\ref{tests} show that the line ratios will be affected by the adopted envelope size and might then give poor fits to multi-transitional line observations if it is systematically over/under-estimated. High angular resolution SiS observations, capable of resolving the emitting regions, exist only for two sources, the carbon stars \object{CW Leo} \citep{Bieging93} and \object{RW LMi}  \citep{Lindqvist00}. In Figs.~\ref{cwleo} and \ref{cit6} we compare our best fit models from Table~\ref{sample} with these interferometric observations. 

For CW Leo the size estimated from Eq.~\ref{eq_size} is 2.2\,$\times$\,10$^{16}$\,cm, which is clearly too large when compared with the interferometric SiS $J$\,$=$\,6\,$\rightarrow$\,5 BIMA observations of \citet{Bieging93} as illustrated in Fig.~\ref{cwleo} 
(solid line). If instead the envelope size is lowered by 23\% to 1.7\,$\times$\,10$^{16}$\,cm a significantly better fit to the observed brightness distribution is found (Fig.~\ref{cwleo}; dash-dotted line). When reducing the envelope size the fractional abundance $f_0$ was increased by 33\% from 1.5\,$\times$\,10$^{-6}$ to 2.0\,$\times$\,10$^{-6}$. This new model also explains the observed multi-transitional single-dish data within 1\,$\sigma$.

In Fig.~\ref{cit6} for RW LMi the model is compared to the Plateu de Bure (PdB) interferometric SiS $J$\,$=$\,5\,$\rightarrow$\,4 line observations by \citet{Lindqvist00} in the ($u,v$)-plane in order to maximize both the sensitivity and the angular resolution. The best-fit model with $r_{\mathrm{e}}$\,$=$\,1.3\,$\times$\,10$^{16}$\,cm (from Eq.~\ref{eq_size}) and $f_0$\,$=$\,4.5\,$\times$\,10$^{-6}$ (Table~\ref{sample}) gives a very good fit to the flux picked up by the PdB interferometer at all baselines (Fig.~\ref{cit6}; solid line). Also shown in Fig.~\ref{cit6} are models with a 30\% smaller envelope size ($r_{\mathrm{e}}$\,$=$\,9.0\,$\times$\,10$^{15}$\,cm and $f_0$\,$=$\,8.5\,$\times$\,10$^{-6}$; dashed line) and 30\% larger envelope size ($r_{\mathrm{e}}$\,$=$\,1.7\,$\times$\,10$^{16}$\,cm and $f_0$\,$=$\,2.8\,$\times$\,10$^{-6}$; dash-dotted line). These models clearly do not reproduce the observations well. All three models are consistent with available single-dish data within 1$\sigma$.

These two examples illustrate the usefulness of angularly resolved emission in further constraining the abundance distribution by lifting the degeneracy between envelope size and fractional abundance that hampers multi-transitional analysis of single-dish data.

Although only two sources have been studied in SiS line emission using interferometers, the fact that both agree within 23\% for the envelope size as estimated by Eq.~\ref{eq_size} gives us confidence in the adopted approach, and we feel that the overall  high $\chi^2_{\mathrm{red}}$ found in the SiS multi-transition analysis are generally not due to this adopted value.

   \begin{figure*}
   \centering{   
   \includegraphics[width=16cm]{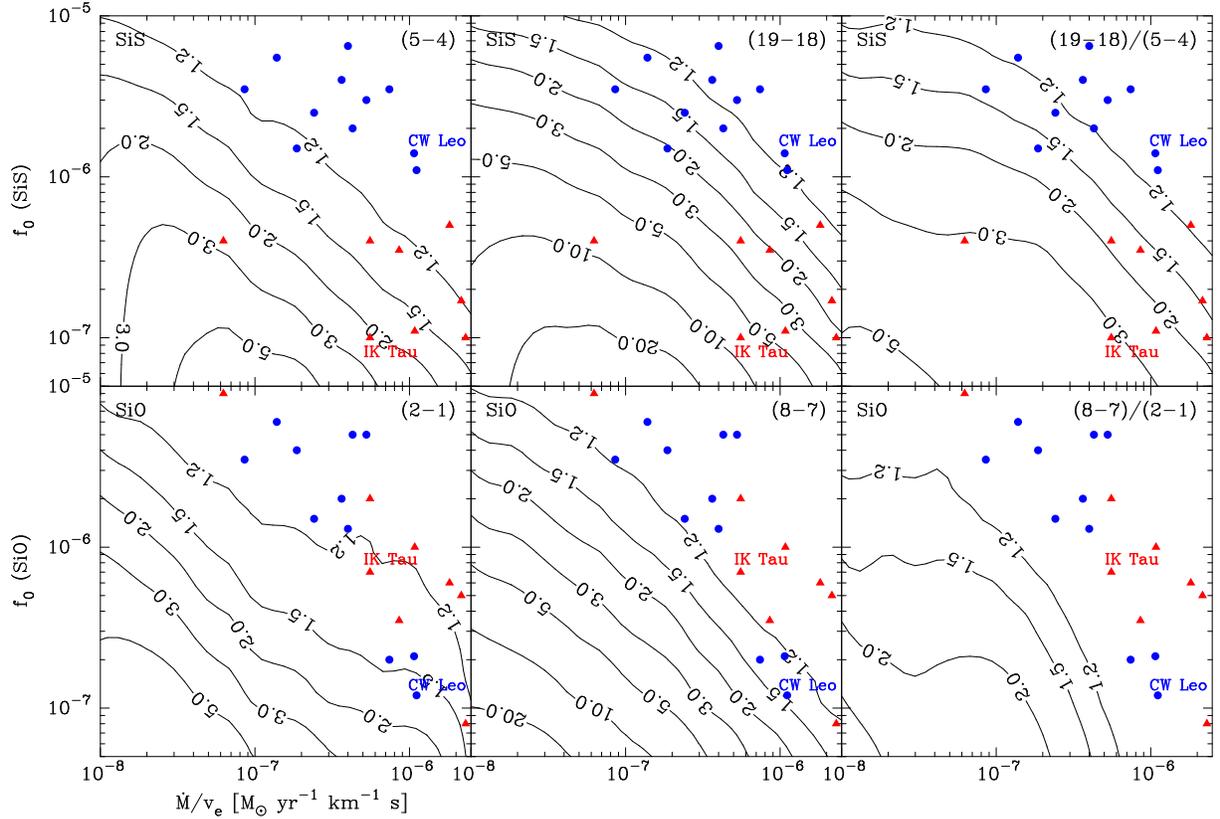}
   \caption{Sensitivity of SiS and SiO circumstellar line emission to the inclusion of an inner compact, high fractional abundance, (pre-condensation) component ($f_{\mathrm{c}}$\,$=$\,2\,$\times$\,10$^{-5}$ and  $r_{\mathrm{c}}$\,$=$\,1\,$\times$\,10$^{15}$\,cm)  as a function of (post-condensation) fractional abundance ($f_0$) and density of the wind ($\dot{M}/v_{\mathrm{e}}$). The contours indicate the ratio of the total line emission for models with and without the compact component (see text for details). Also shown are the fractional abundances ($f_0$) obtained from the excitation analysis for carbon stars (filled circles) and M-type (O-rich) AGB stars (filled triangles) with the two examples CW Leo and IK Tau high-lighted.}
   \label{ratio}}
   \end{figure*}

\subsubsection{A compact, high-abundance, SiS component?}
\label{compact}
From the fact that multi-transitional SiO excitation analyses show a clear correlation of decreasing circumstellar fractional abundance with the density of the wind ($\dot{M}/v_{\mathrm{e}}$) it has been argued that SiO molecules effectively freeze out onto dust grains in the inner part of the winds from AGB stars \citep{Delgado03b, Schoeier06a}. However, it is difficult to put any constraints on the abundance in the pre-condensation region or its spatial size in the analysis of single-dish data as it only contributes very slightly   to the total circumstellar emission \citep{Delgado03b}. Indeed, high angular resolution interferometric observations are required in order to further constrain the abundance profiles and solidify the claim of adsorption of SiO molecules onto dust grains \citep{Schoeier04b, Schoeier06c}.

Of interest is to check whether SiS line emission, as observed by single-dish telescopes in the radio regime, is more sensitive than SiO to such steep chemical gradients.
In Fig.~\ref{ratio} the effect of including a compact region on the line intensity is shown as a function of post-condensation fractional abundance ($f_0$) and density of the wind ($\dot{M}/v_{\mathrm{e}}$). Each panel shows the ratio of line intensities of a model including  also a compact,  pre-condensation, component (with a constant fractional abundance of  $f_{\mathrm{c}}$\,$=$\,2\,$\times$\,10$^{-5}$ out to a cut-off radius $r_{\mathrm{c}}$\,$=$\,1\,$\times$\,10$^{15}$\,cm) to that of a single Gaussian abundance distribution. A grid of models with varying mass loss rates and fractional abundances in the extended component were then produced. The expansion velocity of the wind was taken to be 12\,km\,s$^{-1}$, a typical value found for AGB stars in general \citep{Ramstedt06}. Thermal dust grains were also included in the excitation analysis with $\tau_{10}$\,$=$\,0.1. The star was placed at a distance of 300\,pc and observations were simulated for SiO ($J$\,=\,2\,$\rightarrow$\,1 and $J$\,=\,8\,$\rightarrow$\,7) and SiS ($J$\,=\,5\,$\rightarrow$\,4 and $J$\,=\,19\,$\rightarrow$\,18) line emission using a 15\,m telescope (like the SEST or JCMT). The spatial extents of the SiO and SiS molecular envelopes were calculated using Eq.~\ref{eq_size} for each grid point.

In Fig.~\ref{ratio} also the fractional abundances of SiS and SiO (using a single-Gaussian abundance distribution) found for our sample stars are plotted against the density of the wind ($\dot{M}/v_{\mathrm{e}}$). This is meant as a guide to where in the parameter space the objects are found, and how sensitive the  observed line emission is expected to be to a steep abundance gradient.
From the location of the AGB stars in the panels for SiO in Fig.~\ref{ratio} (botton row of panels) it is clear that any analysis based on single-dish multi-transitional observations is only weakly sensitive to such a compact component consistent with the findings of \citet{Delgado03b}. While the line intensities may be sensitive by up to 50\% to the inclusion of a compact component, the effect on the SiO $J$\,=\,8\,$\rightarrow$\,7/$J$\,=\,2\,$\rightarrow$\,1 line ratio is less than 20\% for all our sample sources. This means that it is generally not possible to separate the compact component from the extended envelope emission from multi-transitional analysis based on single-dish data, and no constraints on the compact component can be found. In addition, the derived post-condensation fractional abundances (using a single-Gaussian abundance profile) may be up to 50\% too high in the worst case (for the majority of sources this effect is less than 20\%). 

In contrast, for SiS it is immediately clear that the results are more sensitive to the inclusion of a compact component (Fig.~\ref{ratio}; top row of panels). This is a combination of the lower opacities in the case of the SiS emission and the fact that the transitions readily observed by current instrumentation in the radio regime probe slightly warmer gas closer to the star. For the lines near 345\,GHz (see Table~\ref{efficiencies}) the SiS $J$\,=\,19 energy level is at 166\,K,  whereas the SiO $J$\,=\,8 level lies at only 75\,K. In particular for the M-type AGB stars, where the circumstellar SiS fractional abundances are low ($\sim$\,3\,$\times$\,10$^{-7}$), the inclusion of a compact high abundance component will drastically affect the observed line ratios and hence the quality of the fits to multi-transitional data  thereby providing an explanation to the high $\chi^2_{\mathrm{red}}$-values reported in Table~\ref{sample}.
This sensitivity to chemical gradients also means that while there is the potential to constrain the compact, pre-condensation, SiS fractional abundance component (if $f_{\mathrm c}$ is sufficiently high), the extended, post-condensation, component will have an added uncertainty. 

From the location of the M-type star IK~Tau in Fig.~\ref{ratio} it is clear that this source may be severely affected by inclusion of a compact, high-abundance component.  Indeed, the single-Gaussian abundance distribution gives a very poor fit to observations with a reduced $\chi^2$ of 9.9 (see Table~\ref{intensities} and Fig.~\ref{model}; solid line). Including a compact component with a constant fractional abundance $f_{\mathrm c}$ of 1.1\,$\times$\,10$^{-5}$ out to a radius $r_{\mathrm{c}}$ of 1.0\,$\times$\,10$^{15}$\,cm, and at the same time lowering the fractional abundance $f_0$ of the extended Gaussian component to 1.0\,$\times$\,10$^{-8}$, significantly improves the fit to the observations as illustrated in Fig.~\ref{model} (dashed line). The reduced $\chi^2$ for IK Tau using the two adjustable parameters $f_{\mathrm c}$ and $f_{\mathrm 0}$ is 1.8. It should also be noted that the fit to the various line shapes is significantly improved (better reproducing the parabolic shapes that are observed), in particular for the high-$J$ transitions where the signal-to-noise is high.  This modelling of IK~Tau is meant to illustrate the fact that a circumstellar model including the effect of adsorption of SiS molecules onto dust grains can explain the observed line intensities. The actual pre- and post-condensation fractional abundances obtained for IK~Tau are uncertain and can not be uniquely constrained from the limited number of observational constraints available. While this procedure could easily be applied to other sources in our sample, thereby improving the fit to the data, more observational constraints are needed in order to make this a meaningful approach.

It has been suggested, based on observations of SiS ro-vibrational transitions in the infra-red, that the carbon star CW Leo (IRC+10216) has a steep SiS fractional abundance gradient in its inner wind \citep{Boyle94}. In their modelling, \citet{Boyle94} find that an initially high fractional abundance of 4.3\,$\times$\,10$^{-5}$, that is decreased by an order of magnitude to 4.3\,$\times$\,10$^{-6}$ at a radial distance of about 8\,$\times$\,10$^{14}$\,cm (at an adopted distance of 200\,pc), best reproduces the data.  
Observations in the radio regime using the Plateu de Bure and BIMA interferometers have also been performed \citep{Bieging93,Lucas95}. \citet{Bieging93} find some indications of a compact SiS component from their interferometric SiS $J$\,$=$\,6\,$\rightarrow$\,5 BIMA observations.  In Fig.~\ref{cwleo} (dashed line) we show that a model with a compact component with a constant fractional abundance $f_{\mathrm c}$ of 1.7\,$\times$\,10$^{-5}$ out to a radius $r_{\mathrm{c}}$ of 5\,$\times$\,10$^{14}$\,cm (corresponding to 8\,$\times$\,10$^{14}$\,cm at 200\,pc) provides a somewhat better fit to the BIMA observations than a model without it (dash-dotted line). These values are also fully consistent with those derived by \citet{Boyle94} when considering their somewhat larger distance and mass-loss rate adopted for CW Leo.  The location of CW~Leo in Fig.~\ref{ratio} suggests that multi-transitional single-dish observations in the radio regime are not expected to be sensitive enough to any SiS abundance gradients. This is further illustrated in Fig.~\ref{model} where it is not possible to discriminate (the difference is less than the calibration uncertainty of 20\%) between a model with a compact component used to fit the BIMA data (dotted line) and a model using a single Gaussian abundance distribution (solid line) with parameters reported in Table~\ref{intensities}.

We note that \citet{Keady93} and \citet{Schoeier06c} found that explaining observed SiO line emission towards CW Leo requires a drastic decrease, by an order of magnitude, in its fractional abundance at a radial distance of about of 5\,$\times$\,10$^{14}$\,cm (for a distance of 120\,pc). Recent high-frequency observations of H$^{13}$CN $J$\,$=$\,8\,$\rightarrow$\,7 line emission towards CW Leo using the SMA suggest that the fractional abundance of HCN remains more or less constant throughout the inner wind and the extended circumstellar envelope \citep{Schoeier07a}. This lends further support to the interpretation that both SiO and SiS have large gradients in their fractional abundances and that this is not the effect of drastically varying physical properties, such as density or temperature, in the envelope.  Additional observations such as these are required for AGB stars in general in order to obtain more reliable constraints on the SiS abundance distribution.

\subsubsection{Time variability}
It has been shown that SiS line emission can be time variable. \citet{Carlstroem90} found that the circumstellar  $J$\,$=$\,5\,$\rightarrow$\,4 and $J$\,$=$\,6\,$\rightarrow$\,5 line emission towards \object{CW Leo} varies on a regular basis that follows the light curve in the K-band of the star. It was also found that when the $J$\,$=$\,5\,$\rightarrow$\,4 line emission was near maximum (also maximum flux in the K-band) the $J$\,$=$\,6\,$\rightarrow$\,5 line emission was near its minimum. The variation in line intensity between maximum and minimum is about a factor of two. Thus observations of these lines at different epochs, which is usually the case in the present analysis, will significantly degrade the quality of the fit of the model to the observations. A plausible explanation for the time variability of certain SiS transitions is overlap of vibrational bands in the infra-red with another molecular species. This will have the effect that the vibrationally excited levels of SiS will be pumped by an additional, but selective, source modifying the excitation in the ground vibrational state. Recent observations of possible maser emission in the SiS $J$\,$=$\,11\,$\rightarrow$\,10,  $J$\,$=$\,14\,$\rightarrow$\,13, $J$\,$=$\,15\,$\rightarrow$\,14 lines by \citet{Fonfria06} towards \object{CW Leo} identify existing overlap of ro-vibrational transitions in the mid-infrared of SiS with other abundant molecules such as C$_2$H$_2$ and HCN as a realistic pumping mechanism.

We note that we find no direct evidence of any maser action in the SiS $J$\,$=$\,20\,$\rightarrow$\,19 line emission observed towards CW~Leo using the APEX telescope. This could mean that this transition is not affected by the selective excitation due to the overlaps or that the relatively large beam of the APEX telescope, compared with the IRAM 30\,m telescope, dilutes the emission from the region very close to the star where the masers are thought to be active. \citet{Fonfria06} derive expansion velocities of about 10\,km\,s$^{-1}$ for their observed lines indicating that they are mainly produced in the acceleration zone ($\approx$\,5\,$-$\,7 stellar radii) before the terminal velocity (14\,km\,s$^{-1}$) of the wind is reached. From modelling of our transitions it is clear that we are not directly probing this region of the wind (see Fig.~\ref{model}). Currently, the effect of time variability due to maser emission is unknown for AGB stars in general.

\section{Circumstellar chemistry}
\label{sect_discussion}
Stellar atmosphere models predict  that the SiS fractional abundance in AGB stars is strongly dependent on the photospheric C/O-ratio (see, e.g., Millar 2003\nocite{Millar03} and Cherchneff 2006\nocite{Cherchneff06}). Typically, SiS fractional abundances under the condition of LTE carbon are expected to be $\sim$\,10$^{-5}$ in carbon stars whereas in M-type AGB-stars they are expected to be only $\sim$\,10$^{-8}$\,$-$\,10$^{-7}$.
From the results reported in Table~\ref{sample} it is clear that the citcumstellar fractional abundance of SiS varies substantially between the carbon star and M-type samples. The mean fractional abundances of SiS found in the present study are in reasonable agreement with LTE stellar atmosphere chemistries. For carbon stars the mean is 3.1\,$\times$\,10$^{-6}$, about an order of magnitude higher than found for the M-type AGB stars (on average 2.7\,$\times$\,10$^{-7}$).

Based on the present modelling, using a single Gaussian abundance profile, there is a weak trend that the SiS fractional abundance for the sample sources  decreases as $\dot{M}$/$v_{\mathrm{e}}$, i.e. the density, in the wind increases (Fig.~\ref{sio_abundance}; top panel). A much stronger correlation is found for the related molecule SiO \citep{Delgado03b, Schoeier06a} as shown in Fig.~\ref{sio_abundance} (middle panel) for the stars also detected in SiS. This has been interpreted as effective adsorption of SiO molecules onto dust grains in the inner wind which has been further corroborated by interferometric observations \citep{Schoeier04b, Schoeier06c}. In addition, there appears to be no way of distinguishing a C-rich chemistry from that of an O-rich based on an estimate of the circumstellar SiO abundance alone, suggesting that non-equilibrium chemical processes are important. For SiS the direct interpretation of the derived circumstellar fractional abundances is more complicated as discussed in Sect.~\ref{high_chi2}, which could explain the weaker correlation of its fractional abundance with the density of the wind.

A departure from LTE is to be expected by the variable nature of AGB stars that induces shock waves that propagates through the photosphere thereby affecting its chemistry. Models of shocked stellar atmospheres \citep{Willacy98, Duari99, Cherchneff06} indicate that the fractional abundances of molecules, such as SiO and SiS, can be significantly altered by the passage of periodic shocks. \citet{Willacy98} in their chemical modelling of CW Leo (IRC+10216) find a thermal equilibrium value for the SiS fractional abundance (relative to H$_2$) of 1.5\,$\times$\,10$^{-5}$ which is increased to 3.4\,$\times$\,10$^{-5}$ due to non-equilibrium chemical processes. For comparison, \citet{Millar03} obtains an LTE SiS fractional abundance of 1.0\,$\times$\,10$^{-5}$. In the present analysis of multi-transitional SiS line emission towards CW~Leo we derive a lower circumstellar value of 1.4\,$\times$\,10$^{-6}$, using a single Gaussian abundance component (see Sect.~\ref{results} and Table~\ref{sample}). However, as discussed in Sect.~\ref{compact} there are observational evidence that the fractional abundance closer to the star may be as high as 1.7\,$\times$\,10$^{-5}$ (see Fig.~\ref{cwleo}) and more in line with the chemical model predictions. The low SiS fractional abundance in the extended, post-condensation, part of the envelope indicates that SiS molecules, like SiO, are significantly incorporated into dust grains.

Similarly, \citet{Duari99} found a fractional SiS abundance (relative to H$_2$) equilibrium value of 4.4\,$\times$\,10$^{-7}$ in their chemical modelling of the M-type AGB star IK Tau which is initially increased by an order of magnitude to 4.8\,$\times$\,10$^{-6}$ near the photosphere due to shocks but is significantly reduced to 8.8\,$\times$\,10$^{-10}$ before the dust formation zone. In comparison, \citet{Cherchneff06} derived a significantly lower LTE value of 7.1\,$\times$\,10$^{-9}$ for the same C/O-ratio of 0.75 and a non-equilibrium value near the photosphere of 1.8\,$\times$\,10$^{-6}$. Also, a drastic decrease of the SiS fractional abundance in the chemically and dynamically active region between 1\,$-$\,5 stellar radii, before dust formation sets in, is found by \citet{Cherchneff06} who derive a very low fractional abundance of $\approx$\,1.0\,$\times$\,10$^{-8}$ for their synthetic model of TX Cam. In our excitation analysis we derive circumstellar SiS fractional abundances of 1.0\,$\times$\,10$^{-7}$ for \object{IK Tau} and 4.0\,$\times$\,10$^{-7}$ for \object{TX Cam}, however, the fits to multi-transitional SiS data are poor, in particular for IK Tau. A significantly better fit is obtained if the fractional abundance close to the star is drastically increased, in the case of IK Tau up to 1.1\,$\times$\,10$^{-5}$ assuming a size of this component of 1.0\,$\times$\,10$^{15}$\,cm (see Sect.~\ref{compact} and Fig.~\ref{model}). Such a high fractional abundance in the inner wind would favour the importance of non-equilibrium processes. Presently, it is not clear if the very low fractional abundance (1.0\,$\times$\,10$^{-8}$) found in the extended envelope around IK Tau is a result of non-equilibrium chemistry in the inner wind or a result of freeze-out onto dust grains. Interferometric observations are clearly required in order to make further progress and better constrain the chemical models. Also the estimated chemical reaction rates involving Si might have to be re-examined \citep{Cherchneff06}.

%Currently there appears to be a discrepancy between measured circumstellar fractional SiS abundances and predictions from photospheric chemical models. In the case of the carbon stars, where model predictions give SiS abundances up to an order of magnitude higher than observed could be the result of SiS becoming incorporated into dust grains. For the M-type AGB stars the chemical models, in particular those including non-equilibrium processes, predict SiS fractional abundances that are an order of magnitude, or more, lower than observationally found. This could indicate that ion-molecule reactions at larger radii, outside the dust formation region, are efficient in producing SiS as suggested by \citet{Scalo80}.

\section{Conclusions}
A systematic and extensive excitation analysis of multi-transition (sub-) millimetre SiS line emission has been performed for a large sample of 19 AGB stars with a range of photospheric C/O-ratios and mass-loss rates. By constraining the circumstellar SiS abundance distribution we reach the following conclusions:

\begin{itemize}

\item In order to derive reliable fractional abundances of SiS, excitation through IR ro-vibrational transitions, needs to be taken into account. This will in most cases (for our sample sources) mean that the effect of dust grains needs to be included in the excitation analysis. The present work contains a proper treatment of this.

\item The adopted SiS envelope size will also affect the quality of the abundance estimates and it has been tested against available interferometric SiS line observations for the two carbon stars \object{CW Leo} (\object{IRC+10216}) and \object{RW LMi} (\object{CIT6}). It is found that the empirical formula derived from SiO observations is adequate to use also for SiS, i.e., these two molecules appear to behave similarly to photodissociation.

\item The mean fractional abundance of SiS in carbon stars of 3.1\,$\times$\,10$^{-6}$ is about an order of magnitude higher than found for the M-type AGB stars (on average 2.7\,$\times$\,10$^{-7}$). Taken at face value, these circumstellar fractional abundances are in reasonable agreement with photospheric LTE chemical models. However, there are indications that the SiS fractional abundance could be significantly higher closer to the star which, at least  in the case of M-type AGB stars, would require non-equilibrium chemical processes.

\item No clear trend that the SiS fractional abundance decreases as the  density ($\dot{M}/v_{\mathrm{e}}$) of the wind increases is found, in contrast to the results for SiO for the same sample of stars. This could mean that SiS molecules are less likely to adsorb onto dust grains than SiO molecules. However, possible time-variability of the line emission in the lower $J$ transitions and the sensitivity of the line emission to abundance gradients in the inner part of the CSE may mask a correlation with the density of the wind. Evidence supporting the presence of steep circumstellar abundance gradients are presented for the carbon star CW~Leo and the M-type AGB star IK~Tau, which can naturally be explained by effective freeze-out of SiS molecules onto dust grains at distances of $\sim$\,5 stellar radii.

\end{itemize}

While it is clear that SiS line emission is a useful tool in investigating the conditions occuring in the chemically and dynamically active region, where dust grains are formed and the stellar wind is accelerated, many uncertainties remain. In particular, the constraints on the detailed SiS abundance distribution are poor. In order to make significant progress, interferometric observations of both low- and high-$J$ SiS line emission in order to more firmly establish the relative importance of freeze-out onto dust grains, photodissociation, and circumstellar chemistry are required. We note that for a typical stellar distance of 300\,pc, next generation interferometers such as ALMA will be required in order to fully resolve the region over which SiS could effectively freeze-out (angular size of $\approx$\,0.2\,$-$\,0.4$\arcsec$).
 
\begin{acknowledgements}
The authors are grateful to the staff at the APEX telescope. The referee, J.H. Bieging, is thanked for insightful comments that helped improve the paper.
% and in particular A. Lundgren. 
FLS and HO acknowledge financial support from the Swedish Research Council.
\end{acknowledgements}

%\bibliographystyle{aa}
%\bibliography{Defs,AGB,Radtransf,Moldata}

\end{document}